\documentstyle[11pt,pasp2,twoside,epsf]{article}
\def\edcomment#1{\iffalse\marginpar{\raggedright\sl#1\/}\else\relax\fi}
\reversemarginpar

\begin{document}
\title{Instabilities and the Formation of Small-Scale Structures in Planetary Nebulae}
 \author{Vikram Dwarkadas}
\affil{Research Centre for Theoretical Astrophysics, School of Physics, A28, Univ of Sydney, NSW 2006, Australia}

\begin{abstract}
	Recent observations have revealed the presence of a variety of
	small-scale ``micro-structures'' in Planetary Nebulae
	images. I discuss numerical hydrodynamics models of
	planetaries and outline the formation and growth of
	instabilities that may be seen in these models. It is possible
	that these may be the forerunners of some of the observed
	small-scale structures.
\end{abstract}
\vspace{-0.3truein}
\section{Introduction}
It has become clear in recent years that the morphology of Planetary
Nebulae (PNe) is much more complicated than was previously
thought. This revolution has been brought about mainly due to
spectacular observations of PNe using the Hubble Space Telescope (see
reviews by Bruce Balick, Howard Bond, Raghvendra Sahai and others in
this volume), accompanied by images of proto-planetary and young
planetary nebulae, especially in infrared bands. The variety of
structure visible at small scales, from rings, arcs, and disks to
FLIERS, Jets and Ansae, is well described elsewhere in this volume. \\

The global morphology of PNe has generally been attributed to the
so-called interacting stellar winds model (Kwok, Purton and Fitzgerald
1978), where a ``fast'' stellar wind from a post-AGB star interacts
with a structured ambient medium, presumably a wind ejected at an
earlier epoch. However the variety of filaments, blobs, fast-moving
knots and ``searchlight beams'' seen in the HST observations are not
easily explained. \\

High-resolution numerical simulations of planetary nebulae reveal the
presence of instabilities that arise during the nebular
evolution. These instabilities are due to various flow patterns, which
in turn depend on the initial conditions and the hydrodynamic model
assumed. It is possible that they may give rise to some of the
small-scale structure visible in PNe images. In this paper we study
the formation and growth of some of these instabilities. More details
can be found in Dwarkadas \& Balick (1998a, hereafter DB98).

\section{Various Types of Models}
Numerical models of Planetary Nebulae can be considered as falling
into these basic categories:

\noindent
{\bf (a) Constant Wind Models} - Models where the wind properties are
constant throughout the evolution of the nebula.
{\bf (b) Evolving Wind Models} - The wind properties change with time during the course of evolution of the nebula.
{\bf (c) Magnetohydrodynamic (MHD) Models} - Magnetic field responsible for the asymmetry.
{\bf (d) Radiation-Hydrodynamic (RHD) Models} - Incorporating
a more realistic treatment of the radiation transfer.\\

In this paper I will discuss predominantly models in the first two
categories. MHD models have been very comprehensively reviewed by
Guillermo Garcia-Segura (this volume), and RHD models by Adam Frank
and Garrelt Mellema (Frank 1999 and references therein).

\subsection{Constant Wind Models}

This is the prototypical PN model, where the very fast ($\approx 1000
\; {\rm km}\; {\rm s}^{-1}$), low density wind collides with the slow
($\approx 10 \;{\rm km}\; {\rm s}^{-1}$), dense wind from a previous
epoch. Since the wind properties are constant, the nebula will reach a
self similar stage in several doubling times of the radius, wherein it
begins to expand with a constant velocity and the shape remains
time-invariant (see for example Dwarkadas, Chevalier \& Blondin 1996,
hereafter DCB96).  The constant expansion velocity does not allow for
the growth of Rayleigh-Taylor instabilities.\\

If an asymmetry exists in the slow ambient wind, with denser material
at the equator compared to the poles, the outcome will be an
aspherical planetary nebula. A pressure gradient is created in the
shocked fast wind due to this asymmetry, with higher pressure at the
equator than the poles. This directs a flow along the walls from the
equator to the poles (Figure 1). In addition, there is a tangential
flow in the opposite direction within the dense shell (DCB96). The
resultant shear flow leads to the growth of Kelvin-Helmholtz (K-H)
instabilities along the edges, giving the edges a rippled appearance,
as is seen in many PNe. The shell appears clumpy, but usually remains
contiguous throughout the simulations. The corrugated appearance is
primarily due to the asymmetry, and spherically symmetric PNe will not
in general be susceptible to this instability.

\begin{figure}
\plotone{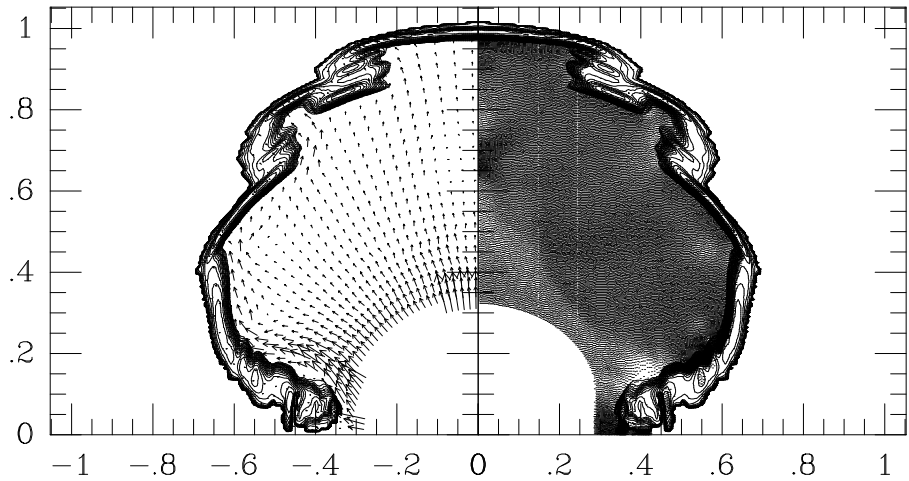}
\caption{Density (RHS) and Velocity Vectors (LHS) from a simulation of
an asymmetrical planetary using a constant-wind model. }
\end{figure}

\subsection{Evolving Wind Models}

Both stellar evolution models and observations of PNe seem to indicate
an evolution in the wind properties over time. Despite significant
progress in these areas (see articles by Schonberner and Wood) we
still do not possess a good theoretical prescription for the
characteristics of the mass loss, telling us how the wind properties
change with time, and the relevant timescales. \\

Here I focus on results of simulations which were performed assuming
that the wind velocity increases as a power law with time.  The wind
mass loss rate decreases correspondingly, such that the mass flux (or
momentum flux) remains constant. The resultant instabilities do not
depend strongly on the manner in which the wind properties vary, and
we concentrate more on the general conditions that lead to the growth
of the instability and in what stage of a PN's development they are
likely to arise.\\

If the velocity and mass loss rate of the so-called ``fast wind'' are
slowly increasing, the nebula will go through an initial
momentum-conserving or radiative stage. A radiative inner shock means
there is no formation of a hot bubble, and the inner and outer shocks
lie very close to each other (DB98), enclosing a thin shell of swept
up material. A shear flow along the inner walls of the asymmetrical
nebula gives rise to K-H instabilities. If the shell is thin these
instabilities may grow to a size comparable to the shell thickness. In
such a scenario the perturbations may continue to grow, giving rise to
the non-linear thin shell instability (NTSI, Figure 2a). This arises
when a thin shell, bounded between two sharp discontinuities, is
perturbed by an amount larger than the thickness of the shell
(Vishniac 1994). The perturbations tend to grow non-linearly, forming
conical projections that are characterised by shear flow within the
unstable fragments (Blondin \& Marks 1998).\\

\begin{figure}
\plottwo{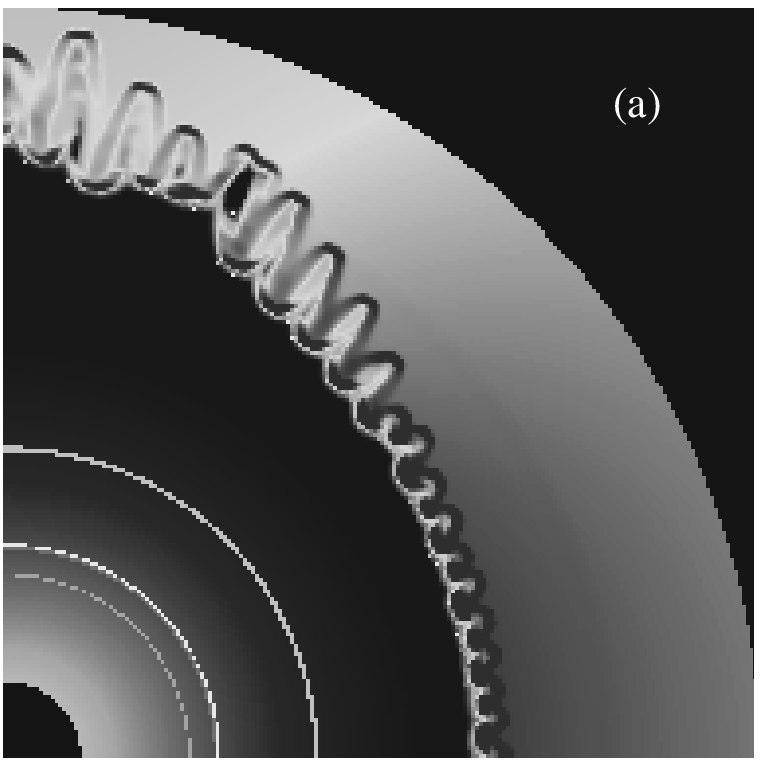}{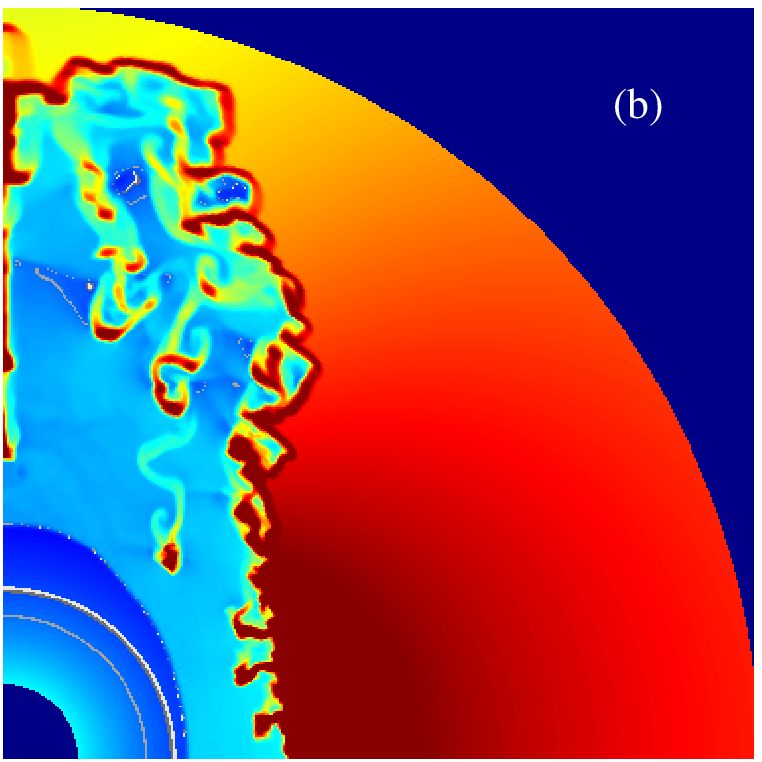}
\caption{(a) The non-linear thin shell instability (b) The R-T
instability.  }
\end{figure}

As the fast wind velocity increases and mass loss rate decreases, the
time taken to radiate away the post-shock energy begins to exceed the
flow rate. The inner shock ceases to be radiative, and a hot bubble of
shocked fast wind separates it from the shocked ambient gas. As long
as the velocity continues to increase, the high-pressure, low density
bubble accelerates the low-pressure, high density swept-up
material. The opposing density and pressure gradients lead to the
growth of the Rayleigh-Taylor (R-T) instability, and wispy filaments
can be seen growing inwards from the edges of the nebula (Figure
2b). These filaments, with bulbous shaped heads caused by the K-H
instability as gas within the nebula flows around them, will continue
to form as long as the wind velocity is evolving. Simultaneously,
there still persists the shear flow along the inner wall of the
nebula. This flow results in some of the filamentary material being
constantly stripped off and mixed into the interior, leading to cold,
dense and slow moving material being mixed in with the
high-temperature gas.  The resultant flow within the hot bubble is
characterised by the formation of vortices and the onset of
turbulence. \\

Herein I have given a flavour for some of the instabilities that may
occur. The filamentary appearance of many nebulae, such as Hubble 5
and NGC 6537, may be due to the R-T instabilities. R-T instabilities
may also occur at ionization fronts (O'Dell \& Burkert 1997). Many
models of asymmetrical planetary nebulae require the presence of a
dense torus of material surrounding the central star. The interaction
of the stellar wind with this torus may lead to the formation of thin
shell instabilities (see Dwarkadas \& Balick 1998b). Other
instabilities may be seen in RHD and MHD simulations. The next step
will be to produce simulated images from the numerical simulations to
enable a proper comparison with the observations.

\acknowledgements I would like to thank the organisers for hosting a
most stimulating and timely conference. Travel support from the
Science Foundation for Physics at the University of Sydney is
gratefully acknowledged.


\end{document}